\documentclass[12pt,aps,preprint]{revtex4}
\usepackage{amsmath}
\usepackage{amssymb}
\usepackage{graphicx}

\begin{document}

\title{Quasi free-standing silicene in a superlattice with hexagonal boron nitride}

\author{T. P. Kaloni, M. Tahir, and U. Schwingenschl\"ogl}

\email{udo.schwingenschlogl@kaust.edu.sa,+966(0)544700080}

\affiliation{Physical Science \& Engineering Division, KAUST, Thuwal 23955-6900,
Kingdom of Saudi Arabia}

\begin{abstract}
We study a superlattice of silicene and hexagonal boron nitride by first principles
calculations and demonstrate that the interaction between the layers of the
superlattice is very small. As a consequence, quasi free-standing silicene is
realized in this superlattice. In particular, the Dirac cone of silicene is preserved, which
has not been possible in any other system so far.
Due to the wide band gap of hexagonal boron nitride, the superlattice
realizes the characteristic physical phenomena of free-standing silicene. In particular,
we address by model calculations the combined effect of the intrinsic spin-orbit coupling
and an external electric field, which induces a transition from a metal to a topological
insulator and further to a band insulator.
\end{abstract}


\maketitle

Graphene is a zero gap semiconductor with very weak spin-orbit coupling (SOC) \cite{geim}. 
Since its discovery a lot of efforts have been undertaken to engineer a finite band gap,
but no satisfactory progress could be achieved. Silicene is closely related to graphene, as
they share the same two-dimensional honeycomb structure, and has been proposed as a
potential candidate for overcoming the limitations of graphene due to its buckled structure
and much stronger SOC. Silicene has first been reported by Takeda and Shiraishi \cite{takeda}
and investigated in more detail in Ref.\ \cite{verri}. While C and Si belong to the same group
in the periodic table, Si has a larger ionic radius, which promotes $sp^3$ hybridization.
The mixture of $sp^2$ and $sp^3$ hybridization in silicene results in a prominent buckling
of 0.46 \AA, which can open an electrically tunable band gap \cite{falko,Ni}. On the other hand,
the band gap induced by the intrinsic SOC was found to amount to 1.6 meV \cite{yao}.
First principles calculations have confirmed that the stable structure of silicene is
buckled \cite{olle}. Similar to graphene, the charge carriers in silicene are expected to behave
like massless Dirac fermions in the $\pi$ and $\pi^*$ bands, which form a Dirac cone at the
K-point. The electronic properties of halogenated and hydrogenated silicene
have been studied by first principles calculations in Refs.\ \cite{houssa,wei} and the effect of 
different substrates on the Dirac cone have been analyzed in Refs.\ \cite{new1,new2,new3}.

Growth of silicene and its derivatives experimentally has been demonstrated for
different metal substrates \cite{padova,vogt,ozaki}. Silicene on a ZrB$_2$ thin film
shows an asymmetric buckling due to the interaction with the substrate, which leads to the
opening of a band gap. However, accurate measurements of the materials properties are difficult
on metallic substrates. In addition, metallic substrates screen externally applied
electric fields and therefore prohibit manipulation of the electronic structure.
For this reason, it would be desirable to achieve free-standing silicene. However,
free-standing silicene probably is instable against a transition into the silicon
structure. A possible solution can be a superlattice that stabilizes the two-dimensional
structure of silicene but still is characterized by a small interaction to the second
component so that the Dirac states are not perturbed. In the following we will substantiate
this idea by first principles calculations. Due to an identical honeycomb structure, the
superlattice of silicene and hexagonal boron nitride
appears to be a promising choice. In addition, hexagonal boron nitride is a wide band gap
semiconductor and therefore makes it possible to study the effects of an external
perpendicular electric field applied to silicene. Because of the remarkably buckled
structure, the intrinsic SOC gap of silicene can be enhanced by a perpendicular external
electric field. Hence, we will study the electronic structure of silicene under an
electric field $E_z$ using band structure calculations as well as an analytical model.

Our calculations are carried out using density functional theory in the generalized gradient 
approximation. Specifically, we employ the Quantum-ESPRESSO package \cite{paolo}. The van der 
Waals interaction \cite{grime,kaloni} as well as the SOC are taken into account. A finite
electric field is applied using the scheme described in Refs.\ \cite{Bengtsson,Meyer}.
The calculations are performed with a plane wave cutoff energy of 816 eV. Furthermore,
a Monkhorst-Pack $8\times8\times1$ k-mesh is employed for optimizing the crystal structure
and a refined $30\times30\times1$ k-mesh is used afterwards to increase the accuracy of
the self-consistency calculation. The supercell employed in our superlattice calculations
comprises one layer of hexagonal boron nitride (18 atoms in a $3\times3$ arrangement) 
and one layer of silicene (8 atoms in a $2\times2$ arrangement). The resulting
lattice mismatch is small (2.8\%) and comparable to that of the frequently studied
superlattice between graphene and hexagonal boron nitride \cite{kaloni,Yankowitz,Giovannetti,dean}. 
We have fully relaxed the lattice parameters of the supercell, finding values of $a=b=7.56$ \AA\
and $c=7.77$ \AA. An energy convergence of 10$^{-8}$ eV and a force convergence of
$4\cdot10^{-4}$ eV/\AA\ are achieved.

\begin{figure*}[ht]
\includegraphics[height=4cm,clip]{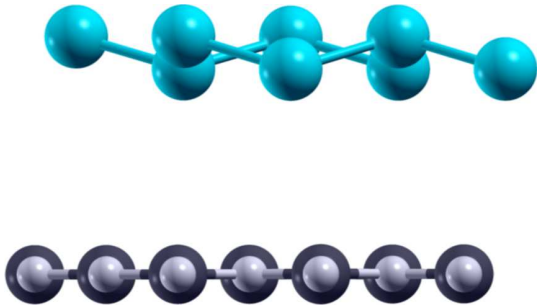}
\caption{Superlattice of silicene (top) and hexagonal boron nitride (bottom) viewed along
the hexagonal $b$-axis.}
\end{figure*}

The structural arrangement of the superlattice under
study is depicted in Fig.\ 1, showing silicene and hexagonal boron nitride layers that
alternate along the $z$-axis. We have also studied superlattices with hexagonal boron nitride
slabs of varying thickness. However, since it turns out that this thickness has
hardly any influence on the silicene electronic states, in particular the charge transfer
between the two component materials, we will focus in the following on the case of one layer
of hexagonal boron nitride alternating with one layer of silicene. Our structural
optimization results in a Si--Si bond length of 2.26 \AA\ and a buckling of 0.54 \AA\ in
the silicene layer. The latter value is slightly but not significantly higher than the
predicted value of free-standing silicene \cite{yao,cheng}. The bond angle between
neighboring Si atoms amounts to 114$^{\circ}$, which agrees well with the value of
116$^{\circ}$ in free-standing silicene. For the interlayer distance between the silicene
and hexagonal boron nitride layers we obtain a value of 3.35 \AA, resembling the 
distance of a silicene layer from $h$-BN \cite{new1,new2,new3}. 

The presence of a Dirac cone has been claimed for silicene grown on metallic substrate
but there is still an ongoing discussion about the validity of this claim
\cite{padova,vogt,ozaki,Lin}. Because of the large band gap of hexagonal boron nitride,
we do not expect B or N states in the vicinity of the Fermi level in the case of our
superlattice, so that the situation is much less involved. The band structure obtained
from our calculations is shown in Fig.\ 2. We observe indeed a well preserved Dirac cone
with a SOC gap of 1.6 meV. Analysis of the partial densities
of states (not shown) clearly demonstrates that the Dirac cone traces back to the $p_z$
orbitals of the Si atoms, while contributions of the B and N atoms are found above 0.6 eV
and below $-1.0$ eV only, with respect to the Fermi energy. We note that the observed
Dirac cone is slightly shifted such that the Dirac point does not fall exactly on the
Fermi energy. It appears at an energy of about 0.04 eV, i.e., the silicene is slightly
hole doped. The energetical shift of the Dirac cone can be attributed to a tiny charge 
transfer between the silicene and the hexagonal boron nitride. Quantitative analysis
shows that the silicene layer loses 0.06 electrons per 8 atoms. However, besides this
small effect (which can be overcome by a minute doping), the charactersitics of
the silicene Dirac cone are perfectly maintained in a superlattice with hexagonal boron
nitride. In the following we will therefore study the effect of an external electric field
on free-standing silicene to describe the properties of the superlattice. In Ref.\ \cite{falko}
the role of the intrinsic SOC and external electric field for the opening of a band gap
have been discussed. The electric field breaks the sublattice symmetry, which induces a finite
band gap. The intrinsic SOC has the same effect. Our calculations (for an ideal buckling
of 0.46 \AA) show that the SOC ($E_z=0$) on its own results in a band gap of 1.6 meV,
which is consistent with the previously reported value in Ref.\ \cite{falko}. To obtain
the same gap by an electric field (without SOC) a value of $E_z=11.2$ meV/\AA\ is needed,
see Fig.\ 3(a). 

\begin{figure*}[ht]
\includegraphics[width=0.6\textwidth,clip]{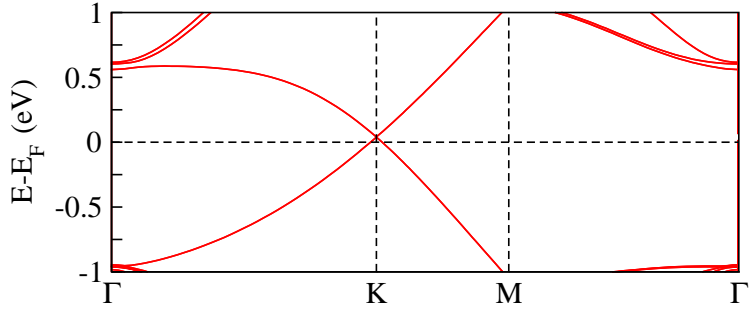}
\caption{Electronic band structure obtained for the superlattice of silicene and hexagonal boron nitride.}
\end{figure*}

From an application point of view, the combined effect of SOC and electric field is of
great interest. We therefore vary $E_z$ relative to the fixed SOC. Band structures
obtained for three different values of the electric field are shown in Figs.\ 3(b) to (d).
For $E_z=3.1$ meV/\AA, see Fig.\ 3(b), we find energy gaps of 1.3 and 7 meV between
the minority and majority spin bands, respectively. When we increase $E_z$ to
3.6 meV/\AA\ the obtained energy gaps change to 1.1 and 9 meV, which we will explain
later by our analytical model. A stronger electric field of $E_z=11.2$ meV/\AA\ leads to
energy gaps of 2.9 and 20 meV. Further enhancement of the electric field results in a
almost linear increase of the energy gaps. The observed dependence of the energy gaps 
on the electric field is much stronger than reported previously \cite{falko,Ni}, because
we take into account the SOC. Our results show that there is no spin degeneracy and a finite 
band gap, which is a combined response of SOC and electric field. In addition,
Figs.\ 3(b) to (d) demonstrate phase transitions from a metal to a topological insulator
and further to a band insulator. The electric field required to obtain a reasonable
band gap is found to be much smaller than typical fields considered before,
which means that the device can be operated in a stable regime at low voltage.

\begin{figure*}[ht]
\includegraphics[width=0.6\textwidth,clip]{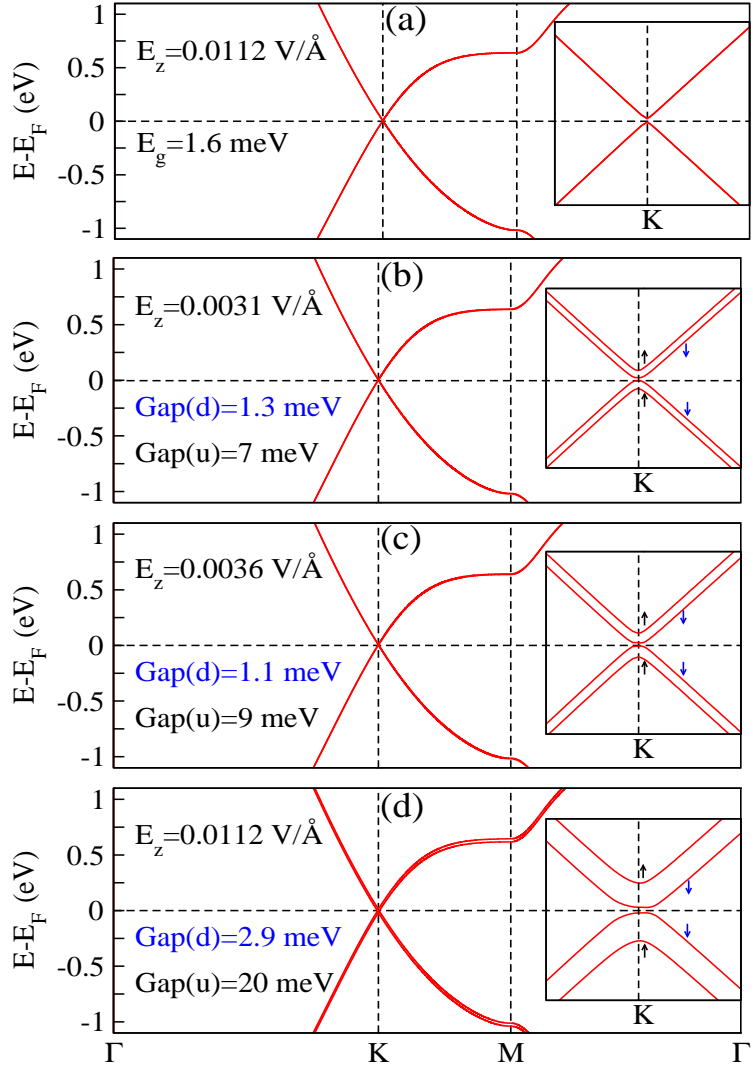}
\caption{Electronic band structure of free-standing silicene: (a) with SOC and
$E_z=0$ or without SOC and $E_z=0.0112$ V/\AA, (b-d) with SOC and different values of $E_z\neq0$.}
\end{figure*}

In order to discuss the mechanisms behind the above observations, we consider an
analytical model. We assume that the silicene sheet lies in the $xy$-plane in 
the presence of intrinsic SOC and an external electric field in $z$-direction. Silicene
can be described by the two-dimensional Dirac-like Hamiltonian
\begin{equation}
H_{s}^{\eta }=v(\eta \sigma _{x}p_{x}+\sigma _{y}p_{y})+\eta s\lambda \sigma
_{z}+\Delta \sigma _{z},  \label{1}
\end{equation}%
where $\eta =+1/{-}1$ denotes the $K$/$K'$ valley, $s=+1/{-}1$ denotes spin up/down,
$\Delta =2lE_{z}$ with $l=0.23$ \AA\ is the electric field, ($\sigma _{x}$, $\sigma _{y}$,
$\sigma _{z}$) is the vector of Pauli matrices, $\lambda $ is the strength of the intrinsic
SOC, and $v$ is the Fermi velocity of the Dirac fermions. For the K valley we have
\begin{equation}
H_{+1}^{K}=v\left( 
\begin{array}{c}
+\lambda +\Delta  \\ 
+p_{x}+ip_{y}%
\end{array}%
\begin{array}{c}
+p_{x}-ip_{y} \\ 
-\lambda -\Delta 
\end{array}%
\right) ,\text{ \ }H_{-1}^{K}=v\left( 
\begin{array}{c}
-\lambda +\Delta  \\ 
+p_{x}+ip_{y}%
\end{array}%
\begin{array}{c}
+p_{x}-ip_{y} \\ 
+\lambda -\Delta 
\end{array}%
\right)   \label{2}
\end{equation}%
and for the K$'$ valley
\begin{equation}
H_{+1}^{K'}=v\left( 
\begin{array}{c}
-\lambda +\Delta  \\ 
-p_{x}+ip_{y}%
\end{array}%
\begin{array}{c}
-p_{x}-ip_{y} \\ 
+\lambda -\Delta 
\end{array}%
\right) ,\text{ \ }H_{-1}^{K'}=v\left( 
\begin{array}{c}
+\lambda +\Delta  \\ 
-p_{x}+ip_{y}%
\end{array}%
\begin{array}{c}
-p_{x}-ip_{y} \\ 
-\lambda -\Delta 
\end{array}%
\right).   \label{3}
\end{equation}%

To obtain the eigenenergies, we diagonalize the Hamiltonian and obtain
\begin{equation}
E_{n,s}^{\eta }=n\sqrt{(v\hslash k)^{2}+(\Delta +\eta s\lambda )^{2}},
\label{4}
\end{equation}%
where $n=+1/{-}1$ denotes the electron/hole band and $k$ is the absolute value of the
wave vector. We next discuss the energy eigenvalues obtained for the K point
to explore the band splitting and quantum phase transitions. The energy gap of
1.6 meV seen in Fig.\ 3(a) as obtained for finite SOC or $E_z$ is consistent with Eq.\ (4),
confirming a metal to insulator transition. Figure 3(b) for finite SOC and $E_z$ with
$\lambda >\Delta=1.4$ meV shows an energy splitting between the spin up and spin down
bands for both the electrons and holes. This splitting is less than the energy gap
between the electrons and holes themselves. In addition, the energy gap between the spin
up bands is greater than that between the spin down bands. This situation reflects a
topological insulating state, which corresponds to the spin polarization regime.
Figure 3(c) is analogous to Fig.\ 3(b) but for
$ \lambda \sim\Delta=1.6 $ meV. We see that the energy gap closes between the 
spin down bands, while the spin up bands maintain a finite energy gap. In the first
principles calculations we cannot reach an exact closure of the spin down gap as
suggested by Eq.\ (4) but obtain a minimum of about 1.1 meV, because of the approximations
involved in the simulations. The situation demonstrated in Fig.\ 3(c)
corresponds to a semi-metallic state. Fig.\ 3(d) is analogous to Figs.\ 3(b) and (c) but
for $ \lambda <\Delta=5.1$ meV. The splitting of the spin down bands has increased as
compared to Fig.\ 3(b), but less than the splitting of the spin up bands. This situation
reflects a band insulator, which corresponds to the valley polarization regime. 
We note that we obtain an identical band structure for the $K'$ point with the spin up
and spin down bands exchanged. The $K$ and $K'$ valleys are non-degenerate due to the
broken inversion symmetry (which is a consequence of the external electric field and the
buckling), compare Eq.\ (4).

In conclusion, we have discussed the structure and electronic properties of a
superlattice of silicene and hexagonal boron nitride. We find that the
Dirac cone of free-standing silicene remains intact in the superlattice due to a
small interaction (the binding energy amount to only 57 meV per atom). A small amount of charge transfer between the silicene and hexagonal
boron nitride results in a slight shift of the Dirac cone towards higher energy, i.e.,
in slight hole doping. Using an analytical model we have analyzed the combined effects
of the intrinsic SOC and an external electric field applied perpendicular to the superlattice.
Our results show that a lifting of the spin and valley degeneracies can be achieved. With
increasing strength of the electric field, the nature of the system changes from a metal
to a topological insulator and further to a band insulator. Therefore, control of the
quantum phase transitions in silicene is possible by tuning the external electric field.

\begin{acknowledgments}
We thank N.\ Singh for fruitful discussions.
\end{acknowledgments}

\end{document}